\documentclass[twocolumn,showpacs,amsmath,amssymb,prl]{revtex4}
\usepackage{graphicx}
\usepackage{dcolumn}
\usepackage{bm}

\begin{document}
\input{epsf}

\title{The Cumulative Bakground of High-Energy Neutrinos from Starburst 
Galaxies}
\author{Abraham Loeb$^{1,2}$ \& Eli Waxman$^3$}
\affiliation{$^1$ Astronomy Department, Harvard University, 60 Garden Street, Cambridge, MA 02138, USA}
\affiliation{$^2$ Einstein Minerva center, Weizmann Institute of Science, Rehovot 76100, Israel,}
\affiliation{$^3$Physics Faculty, Weizmann Institute of Science, Rehovot 76100, Israel}

\begin{abstract}

We show that starburst galaxies convert efficiently cosmic-rays into pions,
which in turn decay into high-energy neutrinos and photons. The cumulative
background of GeV neutrinos is $E_\nu^2\Phi_\nu\approx 10^{-7}~{\rm
GeV~cm^{-2}~s^{-1}~sr^{-1}}$, and its extrapolation to higher neutrino
energies depends on the energy spectrum of the injected cosmic-rays, with
$E_\nu^2\Phi_\nu\propto E_\nu^{-0.15\pm0.1}$ up to $E_\nu\sim0.3$~PeV and
possibly higher neutrino energies. This flux, which constitutes a lower
limit to the high energy extra-Galactic neutrino flux, is potentially
detectable by forthcoming km-scale neutrino telescopes.

\end{abstract}


\date{\today}
\maketitle

Large area, high-energy neutrino telescopes are being constructed to detect
extra-Galactic neutrino sources \cite{HalzenHooper02,Gaisser03}. It is
widely believed that high-energy neutrinos are produced at cosmological
distances based on the fact that the cosmic-ray energy spectrum extends to
$>10^{20}$~eV and is most likely dominated above $\sim3\times10^{18}$~eV by
extra-Galactic protons \cite{Gaisser03}.  High-energy neutrinos are then
likely to be emitted by the same sources that produce the high-energy
protons, through the decay of charged pions generated by interactions of
these protons with the radiation field (or nucleons) within these sources
\cite{Man}.  Gamma-ray bursts (GRBs,\cite{Waxman05}) and jets of active
galactic nuclei (AGN, \cite{Stecker91}) have been suggested as possible
sources of high-energy neutrinos that are associated with high-energy
cosmic-rays.

If the source size is not much larger than the mean-free-path of ultra
high-energy cosmic-rays (UHECRs) for pion production,
as is the case for both AGN jets and GRBs, then UHECR observations set
an upper bound of $E_\nu^2\Phi_\nu<E_\nu^2\Phi_\nu^{\rm
WB}=5\times10^{-8}{\rm GeV~cm^{-2}~s^{-1}~ sr^{-1}}$ on the flux of
high-energy neutrinos produced by pion decay \cite{WBBound}. This
so-called Waxman-Bahcall (WB) upper bound, implies that km (Gton)
scale neutrino detectors are required for detecting the high energy
neutrino flux produced by sources of UHECRs in the energy range
of $\sim1$~TeV to $\sim1$~PeV, and larger detectors are required at
yet higher energies \cite{WBBound}.

The extra-Galactic neutrino number flux per unit energy is expected to be
close to $\Phi_\nu^{\rm WB}$ in the energy range of $10^{19}$--$10^{20}$~eV
\cite{gzk_nu}, since protons originating at cosmological distances with
$>5\times10^{19}$~eV lose all their energy to pion production through their
interaction with cosmic microwave background (CMB) photons \cite{gzk}.
However, the existence of a neutrino flux at the level of $\Phi_\nu^{\rm
WB}$ for lower energies is not guaranteed. Protons in this energy range may
lose only a small fraction of their energy to pion production, making a
neutrino flux much lower than $\Phi_\nu^{\rm WB}$.

In this {\it Letter} we show that radio observations of starburst galaxies
imply a lower limit on their cumulative extra-Galactic neutrino background
flux $\Phi_\nu^{\rm SB}$ which is of order $\Phi_\nu^{\rm WB}$ at the
energy range of $\sim1$~GeV to $\sim300$~TeV (0.3~PeV), and possibly
extending also to higher energies.  Previous discussions \cite{Torres04}
have used a different model-dependent approach to estimate the neutrino
luminosities of individual galaxies and did not integrate the background
neutrino flux over cosmic history.  The main uncertainty in our estimate
involves the spectral index of the injected cosmic-rays. As we show next,
starburst galaxies are "hidden" cosmic-ray sources in the sense that they
dissipate the cosmic-ray energy produced within them, and so their neutrino
background could in principle exceed the WB bound. Nevertheless, the
neutrino flux we derive is comparable to the WB bound, because the energy
production rate of cosmic rays in starbursts is close to the cosmic
production rate of UHECRs. As discussed below, this is not necessarily a
coincidence.

The detection of multi-TeV neutrinos by upcoming km-scale detectors would
provide unique constraints on models for cosmic-ray generation, as well as
a unique handle on the physical properties of the interstellar medium in
starburst galaxies. The expected detection rate may allow km scale
detectors to study the flavor ratio of astrophysical neutrinos, thus
probing neutrino oscillations and physics across unprecedented scales of
energies and lengths \cite{nu_flavors}.

\paragraph*{Starbursts as neutrino factories.}
A substantial fraction of the cosmic star formation activity at redshift
$z\sim 2$ \cite{Reddy,Jun} occurs in transient starburst episodes.  These
episodes are often triggered by galaxy merger events, which channel fresh
gas towards the center of the merger remnant \cite{Mihos,Springel}. The
characteristic scale over which the gas is observed to be concentrated,
($\ell\sim$ hundreds of pc) and the typical gas velocities, ($v\sim$few
hundred ${\rm km~s^{-1}}$) imply a dynamical time of a few million years,
comparable to the lifetime of massive stars. Core-collapse supernovae (SNe)
are therefore expected to enrich the gas with relativistic protons and
electrons, i.e. cosmic-rays, which are accelerated in the collisionless
shocks produced by these explosions. Additional relativistic particles may
be injected by an outflow from a central supermassive black hole, in cases
where quasar activity accompanies the starburst phase
\cite{Genzel,Springel}.  Synchrotron radio emission is routinely observed
in starburst galaxies, confirming the presence of relativistic electrons
within them \cite{Condon01}.

Relativistic protons, injected along with the electrons into the starburst
interstellar medium, would lose energy primarily through pion production by
inelastic collisions with interstellar nucleons. The decay of charged
pions, $\pi^+\rightarrow\mu^++\nu_\mu\rightarrow
e^++\nu_e+\bar{\nu}_\mu+\nu_\mu$ and
$\pi^-\rightarrow\mu^-+\bar{\nu}_\mu\rightarrow
e^-+\bar{\nu}_e+\bar{\nu}_\mu+\nu_\mu$, would then convert part of the
proton energy to neutrinos.  In what follows we estimate the neutrino flux
by showing that: {\it (i)} protons lose essentially all their energy to
pion production, and {\it (ii)} a lower limit to the energy loss rate of
protons may be directly derived from the synchrotron radio flux of the
secondary $e^\pm$.

Protons would lose all their energy to pion production provided that the
energy loss time is shorter than both the starburst lifetime and the
magnetic confinement time within the starburst gas. In the energy range of
interest, the inelastic nuclear collision cross section is $\approx50$~mb,
with inelasticity of $\approx0.5$ \cite{GaisserBook}. The energy loss time,
$\tau_{\rm loss}\approx (0.5 n\sigma_{\rm pp}c)^{-1}$ where $n$ is the
interstellar nucleon density, would be shorter than the starburst lifetime
which is at least a dynamical time, $\sim(2\ell/v)$, as long as
\begin{equation}
\Sigma_{\rm gas}\gtrsim \Sigma_{\rm crit}\equiv {m_p \beta \over \sigma_{\rm
pp}}= 3\times 10^{-2} \beta_{-3}~{\rm g~cm^{-2}}.
\label{sig}
\end{equation}
Here $\Sigma_{\rm gas}\sim m_pn\ell$ is the surface mass density of the
gas and $\beta=v/c=\beta_{-3}(v/300~{\rm km~s^{-1}})$. As it turns out, the
critical surface density, $\Sigma_{\rm crit}$, is comparable to the minimum
$\Sigma_{\rm gas}$ in known starburst galaxies \cite{Thompson05}. 

The ratio of confinement time, $\tau_{\rm conf}$, and loss time, $\tau_{\rm
loss}$, is less straightforward to estimate, since magnetic confinement of
cosmic-rays is not well understood. For starburst galaxies $\Sigma_{\rm
gas}\approx10^9M_\odot/{\rm kpc^2}=0.2{\rm g~cm^{-2}}$, which for a disk
height of $\sim 100$~pc \cite{Condon91} implies $\tau_{\rm
loss}\approx10^5$~yr. The confinement time of 10~GeV protons in our Galaxy
is $\tau_{\rm conf}\simeq10^7$~yr \cite{Yanasak01}. The total gas column
density traversed by protons of energy $\le1$~TeV before they escape the
Galaxy is $\Sigma_{\rm conf}\approx9(E/10{\rm GeV})^{-s}{\rm g~cm^{-2}}$
with $0.5\lesssim s\lesssim0.6$ \cite{Sigma_conf}, suggesting $\tau_{\rm
conf}\simeq10^7(E/10{\rm GeV})^{-0.6}$~yr for the most commonly-used value
of $s=0.6$.  If the confinement time in starburst galaxies was similar to
that of the Galaxy, then the comparison of $\tau_{\rm loss}$ with
$\tau_{\rm conf}$ would have implied that protons with $E\lesssim30$~TeV
lose all their energy prior to their escape from starbursts. However, the
magnetic field in starbursts is much larger than in the Galaxy. Thompson et
al. \cite{Thompson05} have found that the magnetic field strength within
starburst galaxies scales linearly with $\Sigma_{\rm gas}$, and has an
amplitude that is $\sim 100$ times larger than in the Galaxy for
$\Sigma_{\rm gas}=0.2{\rm g~cm^{-2}}$. Assuming that the confinement time
depends on energy only through the proton's Larmor radius ($\propto E/B$),
this suggests that protons with $E\lesssim3\times10^3$~TeV lose all their
energy prior to their escape from typical starburst galaxies.  Moreover,
the neutrino flux is expected to be dominated by starbursts at $z\gtrsim
1$, for which the typical surface density should be even higher than in
local starbursts. It is therefore reasonable to assume that most of the
energy injected into starburst galaxies in $E\lesssim3\times10^3$~TeV
protons is converted to pions.

We now proceed to calibrate the luminosities of starburst galaxies in
high-energy neutrinos based on their observed synchrotron luminosities. As
recently shown by Thompson et al. \cite{Thompson05}, the synchrotron
cooling time of relativistic electrons radiating at a frequency of few GHz
is typically much shorter than the starburst lifetime.  The synchrotron
luminosity reflects, therefore, the energy production rate in these
electrons. The production rate of electrons with energy $E_e$,
$E_e^2d\dot{N}_e/dE_e$, is related to the synchrotron radio luminosity
$L_\nu$ per unit frequency $\nu$ produced by these electrons, as
$E_e^2d\dot{N}/dE_e\approx2\nu L_\nu$. The factor of 2 originates from the
fact that the synchrotron frequency is proportional to the square of the
electron Lorentz factor, $\nu\propto E_e^2$, so that the energy of
electrons in a given decade of $E_e$ is spread over two decades of photon
frequency $\nu$.  Denoting the ratio of injected power in protons and
electron at a fixed particle energy as $\eta_{p/e}$, the luminosity in
$\nu_\mu$ and $\bar{\nu}_\mu$ per logarithmic energy bin $E_\nu$ is given
by $E_\nu dL/dE_\nu=(1/3)\eta_{p/e}E_e^2d\dot{N}/dE_e= (2/3)\eta_{p/e}\nu
L_\nu$. The factor of $1/3$ comes from the fact that $\approx2/3$ of the
proton energy is carried by charged pions ($\sim1/3$ by neutral pions), and
$\approx1/2$ of the pion energy is carried, after it decays, by muon
neutrinos. Cosmic-ray data in the Milky-Way galaxy indicates
$\eta_{p/e}\sim50$ (e.g. \cite{Eichler}). However, we expect a lower ratio
in starburst galaxies. The decay of the charged pions produces secondary
$e^\pm$ which carry $\approx2/3\times1/4=1/6$ of the proton energy (since
the $e^\pm$ carry $\approx 1/4$ of the decaying $\pi^\pm$ energy). This
implies $\eta_{p/e}\approx6$ and a neutrino luminosity $E_\nu
dL/dE_\nu=4\nu L_\nu$. We conservatively ignore the ionization loss by
the electrons \cite{Thompson05}, whose inclusion would only increase our
predicted neutrino flux.

In the Milky way, electrons radiating at few GHz have energies
$E_e\sim10$~GeV. In starburst galaxies the magnetic field is typically
$\sim100$ times larger, implying that the radio luminosity observed is
produced by $E_e\sim1$~GeV electrons.  Since the electrons and neutrinos
produced in pion decays carry similar energy per particle, the radio
luminosity provides a direct estimate of the GeV neutrino luminosity,
$E_\nu dL/dE_\nu(E_\nu=1{\rm GeV})= 4\nu L_\nu|_{\nu\sim1{\rm GHz}}$. The
local 1.4~GHz energy production rate per unit volume is $\nu
(dL_\nu/dV)|_{\nu=1.4\rm GHz} \approx 10^{43}~ {\rm erg~Mpc^{-3}~yr^{-1}}$
\cite{Condon01}. Most of the stars in the Universe formed at redshifts
$z\sim 2$-4 \cite{Ostriker} inside starburst galaxies \cite{Reddy,Jun}.
The fact that the starburst mode of star formation dominated at $z> 2$ is
also supported by the prominence of quasar activity around the same
redshift, as both processes are believed to result from mergers of gas-rich
galaxies.  We extrapolate the local 1.4~GHz energy production rate per unit
volume (of which a dominant fraction is produced in quiescent spiral
galaxies) to the redshifts where most of the stars had formed through the
starburst mode, based on the observed redshift evolution of the cosmic star
formation rate \cite{Ostriker}, and calculate the resulting neutrino
background.  The cumulative GeV neutrino background from starburst galaxies
is then
\begin{eqnarray}
E_\nu^2\Phi_{\nu}(E_\nu&=&1{\rm GeV})\approx \frac{c}{4\pi}\zeta t_H
\left[4\nu (dL_\nu/dV)\right]_{\nu=1.4\rm GHz}
\nonumber\\ 
 &=& 10^{-7}\zeta_{0.5}~ {\rm GeV~cm^{-2}~s^{-1}~sr^{-1}}.
\label{back}
\end{eqnarray}
Here, $t_H$ is the age of the Universe, and the factor
$\zeta=10^{0.5}\zeta_{0.5}$ incorporates a correction due to redshift
evolution of the star formation rate relative to its present-day value.
The value of $\zeta_{0.5}\sim 1$ applies to activity that traces the cosmic
star formation history \cite{WBBound}. Note that flavor oscillations would
convert the pion decay flavor ratio, $\nu_e:\nu_\mu:\nu_\tau=1:2:0$ to
$1:1:1$ \cite{nu_flavors}, so that $\Phi_{\nu_e}=\Phi_{\nu_\mu}
=\Phi_{\nu_\tau}=\Phi_{\nu}/2$.

\begin{figure} [ht]
\centerline{\epsfxsize=3.4in \epsfbox{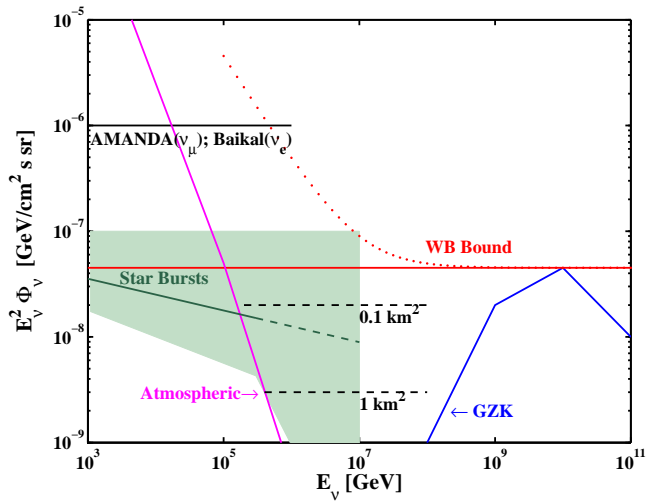}} 
\caption{The shaded region brackets the range of plausible choices for the
spectrum of the neutrino background. Its upper boundary is obtained for a
power-law index $p=2$ of the injected cosmic-rays, and its lower boundary
corresponds to $p=2.25$ for $E_\nu<10^{14.5}$~eV.
The solid green line corresponds to the likely value $p=2.15$ 
(see text). Other lines: the WB upper bound on the
high energy muon neutrino intensity from optically-thin sources;
the neutrino intensity expected from interaction with CMB photons (GZK);
the atmospheric neutrino background; experimental upper bounds of optical 
Cerenkov experiments (BAIKAL \cite{Baikal} and AMANDA \cite{amanda_bound});
and the expected sensitivity
of 0.1~km$^2$ and 1~km$^2$ optical Cerenkov detectors \cite{HalzenHooper02}.}
\label{fig:flux}
\end{figure}

Equation~(\ref{back}) provides an estimate of the GeV neutrino
background. The extrapolation of this background to higher neutrino
energies depends on the energy spectrum of the high energy protons. If the
proton energy distribution follows a power-law, $dN/dE\propto E^{-p}$, then
the neutrino spectrum would be, $E_\nu^2\Phi_{\nu_\mu}\propto E_\nu^{2-p}$.
The energy distribution of cosmic-ray protons measured on Earth follows a
power-law $dN/dE\propto E^{-2.75}$ up to the "knee" in the cosmic-ray
spectrum at a few times $10^{15}$~eV \cite{Eichler,BalloonProton}. (The
proton spectrum becomes steeper, i.e. softer, at higher energies
\cite{Gaisser03}.) Given the energy dependence of the confinement time,
$\propto E^{-s}$ \cite{Sigma_conf}, this implies a production spectrum
$dN/dE\propto E^{-p}$ with $p=2.75-s\approx2.15$.  This power-law index is
close to, but somewhat higher than, the theoretical value $p=2$, which
implies equal energy per logarithmic particle energy bin, obtained for
Fermi acceleration in strong shocks under the test particle approximation
\cite{Krymskii77}. We note that the cosmic-ray spectrum observed on Earth
may not be representative of the cosmic-ray distribution in the Galaxy in
general. The inferred excess relative to model predictions of the $>1$~GeV
photon flux from the inner Galaxy, implies that the cosmic-rays are
generated with a spectral index $p$ smaller than the value $p=2.15$
inferred from the local cosmic-ray distribution, and possibly that the
spectral index of cosmic-rays in the inner Galaxy is smaller than the local
one \cite{innerCR}. The spectrum of electrons accelerated in SNe is
inferred to be a power law with spectral index $p=2.1\pm0.1$ over a wide
range energies, $\sim1$~GeV to $\sim10$~TeV, based on radio, X-ray and TeV
observations (e.g. \cite{SN}).

For a steeply falling proton spectrum such as $dN/dE\sim E^{-2}$, the
production of neutrinos of energy $E_\nu$ is dominated by protons of energy
$E\approx20E_\nu$ \cite{GaisserBook}, so that the cosmic-ray "knee"
corresponds to $E_\nu\sim0.1$~PeV. In analogy with the Galactic injection
parameters of cosmic-rays, we expect the neutrino background to scale as
\begin{eqnarray}
E_\nu^2\Phi_{\nu}^{\rm SB}\approx10^{-7}(E_\nu/1{\rm GeV})^{-0.15\pm0.1}
{\rm GeV~cm^{-2}~s^{-1}~sr^{-1}}
\label{back_h}
\end{eqnarray}
up to $\sim0.1$~PeV. In fact, the "knee" in the proton spectrum for
starburst galaxies may occur at an energy higher than in the Galaxy. The
steepening (softening) of the proton spectrum at the knee may be either due
to a steeper proton production spectrum at higher energies, or a faster
decline with energy for the proton confinement time.  Since both the
acceleration of protons and their confinement depend on the magnetic field,
we expect the "knee" to shift to a higher energy in starbursts, where the
magnetic field is much stronger than the Galactic value. The predicted
neutrino intensity is shown as a solid line in Fig.~\ref{fig:flux}. The
shaded region illustrating the range of uncertainty in the predicted
neutrino background.  This range is bounded from above by the intensity
obtained for $p=2$, corresponding to equal proton energy per logarithmic
bin, and from below by the intensity obtained for $p=2.25$, corresponding
to the lower value of the confinement time spectral index, $s=0.5$.

The extension of the neutrino spectrum to energies $E_\nu>1$~PeV is highly
uncertain.  If the steepening of the proton spectrum at the knee is due to
a rapid decrease in the proton confinement time within the Galaxy rather
than a change in the production spectrum, then the neutrino background may
follow Eq.~(\ref{back_h}) above $1$~PeV, provided that the magnetic
confinement time in starbursts does not decline rapidly at these
energies. This extension is shown as the dashed line in
Fig.~\ref{fig:flux}. If, on the other hand, the steepening at the knee is
due to a steepening in the production spectrum, then the neutrino
background may decrease rapidly following the lower boundary of the shaded
region. Note that although transient sources such as GRBs or AGN may not
contribute at the present-time to the observed Galactic cosmic-rays, they
could dominate the cosmic integral of the neutrino emission by
starburst galaxies \cite{Galactic}. Such a contribution would be
particularly important at $E_\nu\gtrsim0.1$~PeV if the same sources
produce the UHECRs.

\paragraph*{Gamma-ray Production.}
The neutrino emission from $\pi^\pm$ decay will be accompanied by a
comparable flux of $\gamma$-rays from $\pi^0$ decay with energies
$E_\gamma\gtrsim1$GeV. Since the protons carry more energy than the
electrons and the electrons lose only a minor fraction of their energy to
$\gamma$-rays (through inverse-Compton scattering of soft photons
\cite{Thompson05}), the $\gamma$-rays provide a robust measure of the pion
production rate by protons.  This allows a sanity check on our predicted
neutrino fluxes. Our expected $\gamma$-ray flux from the nearest starburst
galaxies must be smaller than existing upper-limits.  For example, analysis
of HESS and EGRET data for NGC253 provided an upper limit of $\sim
1.9\times 10^{-12}~{\rm erg~cm^{-2}~s^{-1}}$ on its $\gamma$-ray flux as a
point source at $E_\gamma\gtrsim 1$~TeV \cite{NGC253,Cillis}. Based on its
1.4GHz flux of $10^{-13}~{\rm erg~cm^{-2}~s^{-1}}$, we predict a
$\gamma$-ray flux of $\lesssim 10^{-12}~{\rm erg~cm^{-2}~s^{-1}}$, well
below the existing upper limit. We have repeated this calculation for other
starburst galaxies with known upper limits \cite{Cillis}, and found similar
results. Our expected $\gamma$-ray fluxes are consistent with previous
models for individual sources \cite{Torres04}, which were based on other
considerations. The forthcoming GLAST
\footnote{http://www-glast.stanford.edu} mission will have the sensitivity
to detect the predicted $\gamma$-ray fluxes from nearby starbursts and
revalidate our expectation for the neutrino background.

\paragraph*{Discussion.} The neutrino intensity predicted by Eq.~(\ref{back_h}) 
at $100$~TeV is $\approx2\times10^{-8\pm0.5}{\rm
Gev~cm^{-2}~s^{-1}\,sr^{-1}}$, implying a detection rate at $E_\nu>100$~TeV
of $\approx10^{1.5\pm0.5}$ events per yr in a 1~km$^2$ detector.  The
neutrino background is therefore potentially detectable by forthcoming
km-scale telescopes. If detected, the signal would serve as a
highly-effective probe of cosmic-ray production in star forming
environments.

As already mentioned, our inferred background of high-energy neutrinos is
similar to $\Phi^{\rm WB}$ since the present-day energy production rate of
cosmic-ray protons by starbursts,
$E^2d\dot{n}/dE=12\nu(dL_\nu/dV)|_{\nu=1.4\rm GHz} \approx 10^{44}{\rm
erg~Mpc^{-3}~yr^{-1}}$, is similar to the energy production rate of UHECRs
\cite{UHECRrate}.  While the neutrino background originates primarily in
starburst galaxies with $\Sigma_{\rm gas}\gtrsim \Sigma_{\rm crit}$,
extra-Galactic cosmic-rays would originate in galaxies with $\Sigma_{\rm
gas}< \Sigma_{\rm crit}$, where they do not encounter substantial loses on
their way out. Since the two classes of galaxies are responsible for making
comparable fractions (up to a factor of $\sim 2$) of the stellar mass
reservoir in the local universe \cite{Jun}, the coincidence between
$\Phi_\nu^{\rm SB}$ and $\Phi_\nu^{\rm WB}$ may simply reflect the matching
condition at $\Sigma_{\rm gas}\sim \Sigma_{\rm crit}$ that known galactic
environments account for the extragalactic flux of UHECRs. Thus, as long as
the sources of UHECRs (such as GRBs or AGN) exist in starburst galaxies,
they would produce a measurable background of neutrinos with energies
$E_\nu\gtrsim 10^{14}$ eV (near the upper envelope of the shaded region in
Fig. 1), that extends well beyond the atomspheric neutrino background.

Our results do not depend on whether the cosmic-rays originate from
supernovae or the supermassive black holes in galactic nuclei.  While
supernova-produced $\gamma$-rays will be spread throughout the star-forming
volume of these nuclei, a black hole origin would bias the $\gamma$--rays
to be centrally concentrated.  GLAST will be able to distinguish between
these possibilities by imaging nearby starburst galaxies in $\gamma$-rays.

\paragraph*{Acknowledgments.}
This work was supported in part by ISF and Minerva grants (E. W.) and by
NASA grants 5-7768 and NNG05GH54G (A. L.).

\end{document}